\def \cth {{\rm coth  }}
\def \sh {{\rm sinh \ }}

\def \ln {{\rm \ ln \  }}

\def \ch {{\rm cosh \ }}
\def \th {{\rm tanh \ }}
\def \const {{\rm const }}
\def \ee {{\rm e}^}

\def \k1 {{1\overk}}

\def \ra {\rightarrow}

\def \a {\alpha}
\def \b {\beta}

\def \l {\lambda}

\def \ep {\epsilon}
\def \r {\rho}

\def \k {\kappa }
\def \d {\delta}

\def \s {\sigma}
\def \t {\theta}

\def \m{\mu}
\def \n{\nu}

\def \e#1 {{\rm e}^{#1}}

\def \1p {{1\over  \pi }}
\def \2p {{{1\over  2\pi }}}
\def \4p {{ {1\over 4 \pi }}}
\def \8p {{{1\over 8 \pi }}}

\newcommand{\rf}[1]{(\ref{#1})}
\newcommand{\beq}{\begin{equation}}
\newcommand{\eeq}{\end{equation}}
\newcommand{\bea}{\begin{eqnarray}}
\newcommand{\eea}{\end{eqnarray}}
\newcommand{\beas}{\begin{eqnarray*}}
\newcommand{\eeas}{\end{eqnarray*}}

\def \del {\partial}
\def \inv {^{-1}}
\documentstyle[titlepage,12pt]{article}
\begin{document}

\begin{tabbing}
\` FIT-HE-94-81\\
\` July 1994\\
\` hep-th/xxx
\end{tabbing}
\addtolength{\baselineskip}{0.2\baselineskip}
\begin{center}
\vspace{36pt}
  {\large \bf Conformally Exact Black Hole Perturbed \\
      by a Marginal Operator}
\end{center}
\vspace{36pt}
\begin{center}
{\bf Kazuo Ghoroku}\footnote{e-mail: gouroku@dontaku.fit.ac.jp}
\end{center}
\vspace{2pt}
\begin{center}
Department of Physics, Fukuoka Institute of Technology,Wajiro, Higashiku,
Fukuoka 811-02, Japan
\end{center}
\vspace{36pt}
\begin{center}
{\bf abstract}
\end{center}
We have examined effective theory induced by
gauged WZW models, in which the tachyon field is added as a marginal
operator.
Due to this operator added,
we must further add the higher order corrections,
which modifies the original configuration, to make the theory
full-conformally invariant. It has been found that
2d is a critical dimension in the sense that the metric obtained from
gauged WZW is modified by the tachyon condensation
for $d>2$, but not for $d\le 2$.

\newpage

\section{Introduction}
Conformal field theories of cosets based on a
gauged WZW model \cite{wit} have provided
several new solutions of string equations. Some solutions have been also
found by
exploiting duality transformations \cite{bus} of the known solutions.
These approaches
opened a new direction for learning more about particles and string theories
in the gravitationally singular configurations, e.g., black hole, black string
and black branes.
However they give semi-classical string solutions, which
are available for $k\rightarrow\infty$,
where $k$ represents the level of a WZW model.

However the conformal invariance of the theory is realized at a finite $k$,
so we need a solution which is
exact to all orders of $1/k$ expansion.
A method to get such exact metric and dilaton has been proposed in
\cite{ant}, \cite{dij}, and the analyses
according to this method have been given in
\cite{bar},\cite{sfet}. The idea is to identify the
Virasoro operator $L_0+{\bar L}_0$ with the Klein-Gordon operator for the
tachyon. Although
this method is intuitive, the exactness of the solutions
was checked perturbatively to three \cite{tseyt1} and four loops \cite{jack}
for the $SL(2,R)/U(1)$ case.
It has also been suggested \cite{tseyt2}\cite{bars}
 how to get the exact solutions
from the functional integral approach.

All the solutions obtained in these ways are however given for
the zero tachyon field. While the tachyon is
expected to be condensed to form a background
configuration from the analysis of the
matrix model and other approach \cite{coop}.
So it is important to examine
the singular background with non-trivial
tachyon configuration in order to get a more realistic insight of particle
and string theories in a sigular configurations.
The purpose of this note is to study
the possibility of a space-time background with singularity like a
black hole with a condensed tachyon field according to a perturbative
method \cite{amb} by starting from a solution of a gauged WZW.
In other words, we could
see the back reaction of the tachyon condensation to the singular
solutions given by gauged WZW models.

\section{Gauged WZW and Perturbation}

A method of deriving the exact
string solutions from gauged WZW
models directly at the $2d$ field theory level has been proposed
\cite{tseyt2}.
Start with the following WZW action
\begin{equation}
S_W(g)=\frac{1}{4\pi}\int_\Sigma Tr(g^{-1}\partial_+gg^{-1}\partial_-g)-
\frac{1}{12\pi}\int_B Tr(g^{-1}dg\wedge g^{-1}dg\wedge g^{-1}dg),
\label{eq:w2}
\end{equation}
where $ g\in G$ and $\partial B=\Sigma$.
By gauging a subgroup $H$ of $G$, we arrive at
the action which is written in a combination of two WZW action.
Then the fully corrected action can be obtained within a WZW.
Then the final
result is obtained in the form
of a non-linear sigma model,
\begin{equation}
      S= \frac{1}{ 4 \pi  } \int d^2 z \sqrt{g}[(G_{\m \n}^{(0)}(X)
           + B_{\m \n}^{(0)}(X))
 (g^{\a\b} + i\ep^{\a\b}) \del_{\a} X^\m \del_{\b} X^\n
    +R\Phi^{(0)}(X) ] \  \ ,
\label{eq:w8}
\end{equation}
where $X^{\m}$ are the invariant elements of $g$.\footnote{
The parametrization of the group elements and which is the invariant
element are shown below.}
However this story seems to be incomplete since unwanted nonlocal terms
appear in the final stage, eq.\rf{eq:w8}.\par
 Then it is convenient to proceed in an alternative way \cite{ant},
where the linear $L_0+{\bar L}_0$ constraint on the tachyon field $T(X)$
is identified with the Klein-Gordon operator,
\beq
  (L_0+{\bar L}_0)T=-\frac{1}{\ee{-2\Phi}\sqrt{G}}
                 \del_{\m}G^{\m\n}\ee{-2\Phi}\sqrt{G}\del_{\n}T,
                       \label{eq:vir}
\eeq
where the Virasoro operators $L_0$ and ${\bar L}_0$ are defined for
the coset $G/H$. Since they are expressed by the differential operators
on the group parameter space, we can read off the exact metric and
the dilaton from eq.\rf{eq:vir}. However the antisymmetric tenser
can not be obtained from this method. But we need not it in the
analysis hereafter.\par
   We notice that
eq.\rf{eq:w8} can be interpreted as a
fully corrected effective action of a 2d gravity which couples to
some renormalizable matter system.
In the case of 2d gravity with conformal matters
\cite{kpz} \cite{ddk}, we know the following solutions,
\beq
  \{G_{\m\n}^{(0)},B_{\m\n}^{(0)},\Phi^{(0)}\}=
           \{\d_{\m\n}, 0, \frac{1}{2}Q_cX^0\}, \label{eq:dd}
\eeq
where $X^0$ is the conformal mode, and the number $Q_c$
is determined by the central charge of the theory.
This configuration is called as the linear dilaton
vacuum.
A perturbative way to search for a more
complicated background is proposed
\cite{amb} by taking an action, which is made of \rf{eq:dd} and
the tachyon part as a marginal operator, as the starting point.
This added tachyon part is corresponding
to the interacting term of the matter fields on the world surface.
We extend this method to the case, where the configuration of
\rf{eq:dd} is replaced by a solution given by a gauged WZW,
in order to investigate the singular configurations which may coexist
with a background tachyon.

In general, the string solutions are obtained by solving the equations
of zero $\beta$-functions of the following action with the
tachyon term $T(X)$,
\beq
 S_{eff}=
{1 \over 4\pi}\int\,d^2z\sqrt{g}
            \left[ {1 \over 2}G_{\mu\nu}(X)g^{\a\b}
              \partial_\a X^{\mu}\partial_\b X^{\nu}
           +R\Phi(X)+T(X) \right], \label{eq:nl1}
\eeq
where $G_{\m\n}$ and $\Phi$ are different from $G_{\m\n}^{(0)}$
and $\Phi^{(0)}$ due to the non-trivial $T$.
The antisymmetric tensor $B_{\m\n}$ has been dropped since
it is not essential here.

Our strategy is as follows; (a):Firstly consider an exact solution
, $G_{\m\n}^{(0)}$ and $\Phi^{(0)}$, which are
obtained from a gauged WZW model for $T=0$. (b):Then solve the
linearized equation of motion for $T$
under the background given in (a). (c):Then solve
the equations for $G_{\m\n}$ and $\Phi$ by using the solution of $T$
given in the step (b), and estimate
their deviations from the original solutions, $G_{\m\n}^{(0)}$
and $\Phi^{(0)}$.\par
 However there is a technical problem in this program.
Since we start from a non-trivial metric $G_{\m\n}^{(0)}$,
which provides the singularity of the curvature,
we need the exact form of equations of $G_{\m\n}$ and $\Phi$
to perform the step (c). But it is impossible to write down them.
So we concentrate our attention on the region far from the singurarity
in order to perform the step (c) in terms of an approximate target
space action given below.
The distance from the singularity is characterized by a mass scale
($\m$) which is related to the black hole mass. In this region
we can investigate the equations by expanding
$G_{\m\n}^{(0)}$ and $\Phi^{(0)}$ around the linear dilaton vacuum \rf{eq:dd}.
\footnote {We can see below that expanding
the solution in terms of $\m$ by assuming its smallness is equivalent
to the expansion of the configuration far away from the singularity.}
Therefore the following approximate target space action
\cite{amb}, \cite{coop}, \cite{das}
is useful in this region,
\beq
S_t=
{1 \over 4\pi}\int\,d^dX\sqrt{G} e^{-2\Phi}
            \bigl[ R-4(\nabla\Phi)^2+
                  {1 \over 16}(\nabla T)^2+{1 \over 16}v(T)-\kappa
            \bigr],
\label{eq:n2}
\eeq
\noindent where the higher derivative terms are suppressed, and
\beq
v(T)=-2T^2+{1 \over 6}T^3+\cdots \  \ . \  \label{eq:n3}
\eeq
$\nabla_\m$ denotes the covariant derivative with
respect to the metric $G_{\m\n}^{(0)}$.
{}From $S_t$, we obtain the following equations,
\bea
\nabla^2T-2\nabla\Phi \nabla T&=&{1 \over 2}v'(T),\label{eq:n4} \\
 \nabla^2\Phi-2(\nabla\Phi)^2&=&-{\kappa \over 2}
                       +{1 \over 32}v(T), \label{eq:n5} \\
 R_{\mu\nu}-{1 \over 2}G_{\mu\nu}R=
            -2\nabla_{\mu}\nabla_{\nu} \Phi &+&G_{\mu\nu}\nabla^2\Phi
            +{1 \over 16}\nabla_{\mu}T\nabla_{\nu}T-{1 \over 32}G_{\mu\nu}
            (\nabla T)^2. \label{eq:n6}
\eea

We must notice that
$G_{\m\n}^{(0)}$ and $\Phi^{(0)}$ do not satisfy
eqs.\rf{eq:n5} and \rf{eq:n6} even if we take $T=0$ since these equations
are the approximate one. However they are correct up to the order of
$O(\m^2)$ when we expand them in power series of $\m^2$,
so we can see the modifications of $G_{\m\n}^{(0)}$ and
$\Phi^{(0)}$ due to $T^{(0)}$ by solving these equations
up to the same order.

\section{d=2; $SL(2,R)/U(1)$ model}

According to the axial gauge invariant formalism,
we can arrive at the well-known 2d
black hole configuration. Here
we take the following parametrization,
\beq
     g=\exp(\frac{i}{2}\t_L\s_2) \exp(\frac{1}{2}r\s_1)
            \exp(\frac{i}{2}\t_R\s_2),   \label{eq:b1}
\eeq
and $X^{\m}=(r,\t)$, where $\t\equiv\frac{1}{2}(\t_L-\t_R)$.
Then we can get the following configurations,
\bea
 G_{\mu\nu}^{(0)} &=& \frac{k-2}{2}{\rm diag}(1,\b(r)), \ \ \ \
             \b(r)=\frac{4}{\cth^2(\frac{r}{2})-\frac{2}{k}}
                                \label{eq:b3}  \\
      -2\Phi^{(0)}&=&{\rm a}+\frac{1}{2}\ln [\sh^2(r)/\b(r)]
                   =\const +\ln(\sh r)-\frac{1}{2}\ln G^{(0)}, \label{eq:b4}
\eea
where $G^{(0)}={\rm det}G_{\m\n}^{(0)}$ and
${\rm a}$ is a constant.

It is instructive to expand this configuration by the black hole mass,
$M_b$, by assuming its smallness.
$M_b$ is introduced here according to \cite{wit},
\beq
   -2\Phi^{(0)}(r=0)={\rm a}=\ln (\frac{1}{2}QM_b). \label{eq:b8}
\eeq
By the reparametrization,
\beq
   -2\Phi^{(0)}={\rm a}+\frac{1}{2}\ln [\sh^2(r)/\b(r)]
             =\frac{1}{2}Q\r. \label{eq:b6}
\eeq
and $\t'=\sqrt{k'/k}\t/Q$, the metric $G_{\m\n}^{(0)}$ is rewritten
for $M_b\ll 1$ and $\r\geq 0$ as follows,
\beq
    ds^2=
    \frac{1}{2}kQ^2\bigl( \frac{d\r^2}{{\bar f}(\r)}
           +{\bar f}(\r)d\t'^2 \bigr),
\ \  {\bar f}(\r)=\ee{-Q\r}\sh^2(r)\approx
             4\frac{k}{k'}(1-{\bar \m}\ee{-\frac{Q}{2}\r}),
\label{eq:b10}
\eeq
where $2{\bar \m}=QM_b\sqrt{k/k'}$. We can see that
\rf{eq:b10} is equivalent to an approximate black hole solution
found in \cite{msw}.

This $M_b$ expansion is corresponding to
expanding the solutions in terms of the series of
$\ee{-r_0}\equiv\m^2$ with the replacement of $r$ by $r+r_0$.
The parameter $r_0$ is defined by \rf{eq:b6} at $\r=0$ as,
\beq
   {\rm a}+\frac{1}{2}\ln [\sh^2(r_0)/\b(r_0)]
             =0. \label{eq:b116}
\eeq
{}From this and \rf{eq:b8}, we can see $M_b\propto \ee{-r_0}$ for $r_0>>1$.
Then $M_b$ expansion is corresponding to examining
the equations at
large $r(>r_0)$, where the space-time is asymptotically flat but
the tail of the black hole can be seen.

{}From \rf{eq:n3} and \rf{eq:n4}, the linearized
equation of $T^{(0)}$ is written as
\beq
  [\del_r^2+\frac{1}{2}\del_r(\ln G^{(0)})\del_r+\del_r(-2\Phi^{(0)})\del_r
           +k-2]T^{(0)}=0, \label{eq:b110}
\eeq
by assuming as $T^{(0)}=T^{(0)}(r)$. This linearization is justified
by introducing a small parameter $\l$ as given below.
Eq.\rf{eq:b110} can be rewritten as,
\begin{equation}
z(1-z)\frac{d^2T^{(0)}}{dz^2}+(c-(a+b+1)z)\frac{dT^{(0)}}{dz}-ab
T^{(0)}=0, \label{eq:b12}
\end{equation}
where $z=\ch^2(r)$ and
\beq
a=\frac{1}{4}(1+\sqrt{9-4k}), \ \
b=\frac{1}{4}(1-\sqrt{9-4k}), \ \  c=\frac{1}{2}. \label{eq:b13}
\eeq
The solution of Eq.\rf{eq:b12} is known as the hypergeometric
function, $F(a,b;c;z)=F(b,a;c;z)$.
We firstly solve Eq.\rf{eq:b12} for $d<2$ ($k<9/4$),
then we take the limit of $d=2$ ($k=9/4$) of the solution.
For large $r$ (or $z$), it is convenient to consider the following
two independent solutions,
\beq
       w_1= z^{-a}F(a,a-c+1,a-b+1,z\inv), \   \
       w_2= z^{-b}F(b,b-c+1,b-a+1,z\inv). \label{eq:b15}
\eeq

Generally we should take a linear combination of
$w_1$ and $w_2$. However, to fix the ratio between them is difficult.
Since $w_1$ and $w_2$ in $T^{(0)}$ are corresponding to two
independent interaction terms on the world surface,
the problem of fixing the ratio of them
is similar to the "Big Fix" \cite{cole}
proposed in the 4d quantum gravity in terms of the wormhole interaction.
So this might be solved by considering the topology changing
interactions of 2d surface.
But it is out of our present task. While
$w_2$ is favourable if we demand that
$T^{(0)}$ should coincide with the one obtained in \cite{ddk} in the
limit of $r\ra \infty$.\footnote
{In this limit, the equation for
$T^{(0)}$ is written as $[\hbar\del_r^2+\del_r+k-2]T^{(0)}=0$, where
$\hbar$ is maintained. This equation is solved by assuming the form,
$T^{(0)}=\ee{-\a r}$, and we must choose the solution
$\a=(1-\sqrt{1-4(k-2)\hbar})/2\hbar$ in order to get from it the
classical limit, $\a=k-2$, for $\hbar\ra 0$.}
However $w_1$ can be neglected at large $r$ compared to $w_2$
even if we retain $w_1$ through an appropriate linear combination
of $w_1$ and $w_2$. So we consider
\[ T^{(0)}=w_1+w_2 , \]
as a solution in order to extend this for $d>2$.
\vskip 3pt
   Nextly, we solve the approximate equations of $G_{\m\n}$
and $\Phi$ in terms of $T^{(0)}$ and the following parametrization
\cite{amb},
\bea
G_{\mu\nu}&=& G_{\m\n}^{(0)}+\lambda^2h_{\mu\nu}+\cdots ,
           \  \  \Phi=\Phi^{(0)}+\lambda^2\Phi^{(2)}+\cdots,
           \nonumber \\
     & & \ \ {} T=\lambda(T^{(0)} +\lambda T^{(1)}+\cdots), \label{eq:b16}
\eea
where $h_{\m\n}={\rm diag}(0,h(r))$ and $\l$ represents a small parameter.
After solving the equations of order $\l^2$ of \rf{eq:n5} and \rf{eq:n6}
by expanding $G_{\m\n}^{(0)}$, $\Phi^{(0)}$ and $T^{(0)}$ in the power
series of $\m^2$ up to $O(\m^2)$, we obtain
\beq
  \Phi^{(2)}=\frac{\m^2}{128}\ee{Q_0(1-\ep)r}, \ \
  h=0. \label{eq:b22}
\eeq
where
\beq
 Q_0=2\del_0\Phi^{(0)}\mid_{\m^2=0}=-\sqrt{\frac{2}{k'}}, \ \
        \ {} \ep=\sqrt{9-4k}. \label{eq:b21}
\eeq

Then the 2d solution is obtained by taking the limit of
$k=9/4$ and $\ep=0$. However the
results are useful for $d<2$ (but not for $d>2$),
and the second result, $h=0$,
is independent on $\ep$ namely on $d$.
This implies that
the 2d black hole configuration does not get any modification from
the tachyon condensation. While the dilaton is modified. Then a
black hole could coexist with the condensed tachyon as
an exact string solution. Nextly, we consider the case of $d>2$ in the
following section.

\vskip 3pt
\section{d$>$2; $SL(2,R)\otimes SO(1,1)^{d-2}/SO(1,1)$ model}
\vskip 3pt
 We consider the model $SL(2,R)$ $\otimes$ $SO(1,1)^{d-2}/SO(1,1)$ as
an example of d- dimensional target space.
According to \cite{gin}, we parametrize
the group element of $SL(2,R)\otimes SO(1,1)^{d-2}$ as,
\beq
 g=\pmatrix{g_0 & 0 & \ldots & 0 \cr
            0   & g_1 & \ldots & 0 \cr
            \vdots & \vdots & \ddots & \vdots \cr
            0 & 0 & \ldots & g_{d-2} \cr}, \  \  {}
   g_0=\pmatrix{p & u \cr
            -v & q \cr}, \ \  pq+uv=1, \label{eq:b24}
\eeq
where $g_0$ represents the $SL(2,R)$ part, and
for the each $SU(1,1)$,
\beq
   g_i=\pmatrix{\ch t_i & \sh t_i \cr
                \sh t_i & \ch t_i \cr}, \ \  i=1\sim d-2. \label{eq:b25}
\eeq
The more convenient parametrizations are given after the gaugings
are fixed.
The embedding of the subgroup $H=SO(1,1)$ in $G$ is chosen as,
\beq
  \s=\pmatrix{s_0 & 0 & \ldots & 0 \cr
            0   & s_1 & \ldots & 0 \cr
            \vdots & \vdots & \ddots & \vdots \cr
            0 & 0 & \ldots & s_{d-2} \cr}, \  \
   s_0=e_0\pmatrix{1 & 0 \cr
            0 & 1 \cr}, \ \
   s_i=e_i\pmatrix{0 & 1 \cr
             1 & 0 \cr},  \label{eq:b27}
\eeq
with the normalization of the coefficients, $\Sigma_{i=0}^{d-2}e_i^2=1$.
\vskip 3pt
 In this case, the original action is characterized by (d-2) $SO(1,1)$
WZW actions $S_W(g_i)$ of level $k_i$ and the previous gauged WZW
action of level $k$.
The procedures to obtain the exact metrics are parallel to
the previous section.

We firstly consider the case of the vector gauging.
Under the vector gauge transformation,
$g\ra u^{-1}gu$ and $u\in H$, we can see
$\d p=\d q=\d t_i=0$. This implies that the ground state is
independent on $u$ and $v$. The resulting metric and the dilaton are
separately given for two cases, (i)$pq>0$ and (ii)$pq<0$.
For the case (i), $G_{\m\n}^{(0)}$ and $\Phi^{(0)}$ are given as,
\bea
         ds^2 &=&\frac{k-2}{2}\bigl(dR^2-\frac{dX_0^2}{\th^2R-2/k}
                  +\sum_{i=1}^{d-2}dX_i^2\bigr), \label{eq:b28} \\
         -2\Phi^{(0)}&=&\const +\ln\sh 2R-\ln\sqrt{-G^{(0)}}, \label{eq:b29}
\eea
where $G^{(0)}={\rm det}G_{\m\n}^{(0)}$.
The $SL(2,R)\otimes SO(1,1)^{d-2}$ variables are reparametrized as
follows;
\beq
 p=\ch R\ee{X_0}, \ \ q=\ch R\ee{-X_0}, \label{eq:b30}
\eeq
and $X_i$ ($i=1\sim d-2$) are the linear combinations of $t_i$
\cite{gin}. In the case of (ii), we obtain the same form with \rf{eq:b28}
where $\th R$ is replaced by $\cth R$
with the parametrization,
\beq
 p=\sh R\ee{X_0}, \ \ q=-\sh R\ee{-X_0}. \label{eq:b32}
\eeq
The eq.\rf{eq:b29} is common to two cases (i) and (ii) but with
different $G^{(0)}$'s.

 For the axial vector gauge invariant formalism,
the $SL(2,R)$ parameters, $u$ and $v$, are unchanged under the axial
gauge transformation,
$g\ra ugu$,
then the ground state is expressed in terms of
$u,v$ and $t_i$ (namely $X_i$).
We consider the two kinds of metrics in this case also, for
(i)$uv>0$ and for (ii)$uv<0$. For $uv>0$,
\bea
         ds^2 &=&\frac{k-2}{2}\biggl(dR^2
            -\frac{dX_0^2}{(1+\eta)\th^2R-\eta-2/k} \nonumber \\
             & & \ \ \ \ {} +\frac{dX_{d-2}^2}{1+\eta-2/k-\eta\cth^2R}
                  +\sum_{i=1}^{d-3}dX_i^2\biggr), \label{eq:b33}
\eea
in terms of the parametrization \rf{eq:b30}. Here
$\eta=\sum_{i=1}^{d-2}(e_i/e_0)^2(k_i/k)$. And the metric for $uv<0$
is given by replacing $\th R$ by $\cth R$ in \rf{eq:b33} with
the parametrization \rf{eq:b32}, and
$\Phi$ has the same form with \rf{eq:b29} for both the cases with
each $G^{(0)}$.

To perform the same analysis done in section {\bf 3},
we rotate $X_0$ as $X_0\ra i\t$ for the sake of the unitarity.
By changing the variable
as $2R\equiv r$ and assuming that $T^{(0)}$ depends on $r$ only,
we arrive at the same
linealized equation of the tachyon with \rf{eq:b12}
for the above four cases in spite of the
difference of $G_{\m\n}^{(0)}$ and the dimension.
While $\ep$ becomes pure
imaginary here since $(26>)d>2$ and
\beq
  9-4k=\frac{2-d}{26-d}, \label{eq:b35}
\eeq
which is the condition of the conformal invariance.
Then $a$ and $b$ given in \rf{eq:b13} are complex
conjugate each other.
Here we restrict our attention to the real solution since we can not
give a physical meaning to the complex solutions.
So we choose the same $T^{(0)}$ with that given in
section {\bf 3},
\beq
  T^{(0)}=w_1+w_2\sim 2\sqrt{2}\m\ee{\frac{1}{2}Q_0r'}
           \cos (\frac{\d}{2}Q_0r'+\a), \label{eq:b42}
\eeq
where $\d=\sqrt{4k-9}$ and $r'=r-r_0$. The second expression is the
$\m$ expansion stated before. \par
 Next, we solve the equations of other fields in the large $r(>r_0)$
region, where $r_0$ is defined by \rf{eq:b116}, by expanding
$G_{\m\n}^{(0)}$ and $\Phi^{(0)}$ in the series of $\mu^2$ as in the
previous section. Here we set the coordinates as
$X^{\m}=(r,\t,X^{d-2},X^i)$, $i=1\sim d-3$, $\m=0,1,2, \ldots, d-1$, and
use eq.\rf{eq:b16} with the following parametrization,
\beq
   h_{\m\n}={\rm diag}(1,h_1,h_2,\vec{0}), \label{eq:b36}
\eeq
where we take $h_2=0$ for vector gauging. After a calculation
we obtain the following result which is common to the above four
gauged WZW models,
\bea
       \Phi^{(2)}&=&\biggl(
          \frac{\m^2}{256}\cos(\d Q_0r'+2\a)+\phi_c\biggr)
              \ee{Q_0r'},     \label{eq:b44} \\
       (h_1+h_2)&=&\frac{4}{Q_0^2}
                   \bigl(Q_0^2\phi_c-\frac{\m^2}{32}\bigr)
                   \ee{Q_0r'}, \label{eq:ac41}
\eea
where $\phi_c$ is a constant which can not be determined by
solving the equations.

 In order to decide $\phi_c$, we require a reasonable condition that
$\Phi^{(2)}$ should coincide with that of 2d (\rf{eq:b22})
when we take the limit
$d\ra 2$ in eq.\rf{eq:b44}. From this requirement, we get
\beq
   \phi_c=\frac{\m^2}{256}. \label{eq:bc41}
\eeq
Except for the undetermined ratio of $h_1$ and $h_2$, we finally
obtain the following
results,
\beq
       \Phi^{(2)}=\frac{\m^2}{256}[1+\cos(\d Q_0r'+2\a)]\ee{Q_0r'},
                    \hspace{.5cm}  \ \
       (h_1+h_2)=-\frac{\m^2}{64}
                  \frac{d-2}{26-d}\ee{Q_0r'}. \label{eq:ab42}
\eeq
We notice that $h_1+h_2$ vanishes at $d=2$ limit, and this is
consistent with the 2d result \rf{eq:b22}. \par
In order to continue \rf{eq:ab42} to $d\le 2$, we should consider
the case of vector gauging where we could take as $h_1\equiv h$ and
$h_2=0$. The previous result \rf{eq:b22}
shows that $h=0$ for $d\le 2$, but \rf{eq:ab42}
does not lead to this result. So we can not apply \rf{eq:ab42} for $d<2$.
The same thing is said for $\Phi^{(0)}$ also. This implies that
there are different phases above and below the critical dimension $d=2$.
We should also notice the next curious
point. For $d>2$,
both corrections $h$ and $\Phi^{(2)}$ appear, but the behavior of
damping osccilation of $T^{(0)}$ with $r$ is reflected
on $\Phi^{(2)}$ only and
not on $h$. \par
The result of non-zero $h_1+h_2$ implies
that the tachyon condensation disturbes the singular metric
configurations for $d>2$. The metric $G_{\m\n}^{(0)}$'s are expanded by
$\m^2$ as follows,
\beq
  G_{\mu\nu}^{(0)}={\rm diag}
             (1,1\pm4 \m^2\ee{Q_0r'},0,{\vec 0}\ \ )+O(\m^4)
                    , \label{eq:ab43}
\eeq
for the vector gauge, and
\beq
  G_{\mu\nu}^{(0)}={\rm diag}
             (1,1\pm 4\m^2(1+\eta)\ee{Q_0r'},
              1\pm 4\m^2\eta\ee{Q_0r'},{\vec 0})+O(\m^4)
                    , \label{eq:ab44}
\eeq
for the axial vector gauge. The upper (lower) sings are obtained for the
parametrization \rf{eq:b30} (\rf{eq:b32}). Then we can say that
the tail of the original, singular metric, $G_{\m\n}^{(0)}$, of $O(\m^2)$
could be canceled out by $h_i$
by choosing an appropriate value of $\l$.
Even if we could not cancel out
the higher order terms of $\m^2$ in $G_{\m\n}^{(0)}$ by the
higher order corrections due to tachyon condensation, the final form of
the modified $G_{\m\n}$ could not maintain the original structure
of $G_{\m\n}^{(0)}$, i.e. the black hole (string) structure. In this sense,
the singularity of space-time seems to be eliminated
through the tachyon condensation for $d>2$, especially at
the realistic dimension $d=4$. This point may be an important result.
However we need more higher order calculations to
assure this point. We will discuss on this point elsewhere.

\section{Conclusion}

 Here we find that
the 2d black hole is not affected by the condensation of
the tachyon, at least up to the quadratic order of the tachyon field.
On the other hand, the dilaton gets a modification.
However for $d>2$, both the metric and the dilaton
are modified,
and these modifications are coinciding with that of 2d case at the limit
of $d=2$. However the solutions obtained for $d>2$ can not be continued
to the region of $d<2$ and vice versa. Then we might say that
a kind of phase transition, where the order parameter is the shift
of the metric from the WZW solution,
has happened at the critical dimension $d=2$.
This implies that the singularities of the space-time manifold
could be removed by considering the tachyon condensed vacuum for
$d>2$.


\newpage
\begin{thebibliography}{99}

\bibitem{wit}E. Witten, Phys. Rev. D44 (1991) 314.
\bibitem{bus}T.Buscher, Phys. Lett. B201(1988)466;ibid B194(1987)59.
\bibitem{ant} I.Antoniadis, C.Bachas, J.Ellis and D.V.Nanopoulos
          Phys. Lett. B211 (1988) 393.
\bibitem{dij}R. Dijkgraff, H. Verlinde and E. Verlinde, nucl. Phys. B371
             (1992) 269.
\bibitem{bar}I.Bars and K.Sfetsos, Phys. Rev. D46(1992)4501;Phys. Lett.
             B301(1993)183.
\bibitem{sfet}K. Sfetsos, Nucl. Phys. B389 (1993) 424.
\bibitem{tseyt1}A.A.Tseytlin, Phys. Lett. B268 (1991) 175.
\bibitem{jack}I.Jack, D.R.Jones and J.Panvel, Nucl. Phys. B393 (1993) 95.
\bibitem{tseyt2}A.A.Tseytlin, Nucl. Phys. B399 (1993) 601.
\bibitem{bars}I.Bars and K.Sfetsos, Phys. Rev. D48(1993)601.
\bibitem{coop}A.Cooper, L.Susskind and L.Thorlacius,
             Nucl. Phys. B363 (1991) 132.
\bibitem{amb}J.Ambjorn and K.Ghoroku, NBI preprint NBI-HE-93-63.
\bibitem{kpz}V. Knizhnik, A. Polyakov and A. Zamolodchikov, Mod.Phys.Lett
             A3 (1988) 819.
\bibitem{ddk} F. David, Mod.Phys.Lett. A3 (1988) 1651; J. Distler
              and H. Kawai, Nucl.Phys. B321 (1989) 509.
\bibitem{msw}G.Mandal, A.M.Sengupta and S.R.Wadia, Mod. Phys.
 Journ. A (1991)1685.
\bibitem{das}S. R. Das and  B.Sathiapalan, Phys. Rev. Lett.
                    56 (1986) 2664;
             C.Itoi and Y.Watabiki, Phys. Lett. B198 (1987) 486; A.
              A. Tseytlin,  Phys. Lett. B264 (1991) 311.
\bibitem{cole}S. Coleman, Nucl. Phys. B310(1988)643.
\bibitem{gin}P. Ginsparg and F. Quevedo, Nucl. Phys. B385 (1992) 527.

\end{thebibliography}
\end{document}